\documentclass[twocolumn,twocolappendix]{aastex631}
\shorttitle{Spatio-temporal characterization of Cassiopeia~A}
\shortauthors{Y. Ichinohe and T. Sato}
\graphicspath{{./}{figures/}}

\begin{document}

\title{Spatio-temporal characterization of Cassiopeia~A}

\author[0000-0002-6102-1441]{Yuto Ichinohe}

\author[0000-0001-9267-1693]{Toshiki Sato}
\affiliation{Rikkyo University \\
3-34-1 Nishi-Ikebukuro, Toshima-ku \\
 Tokyo 171-8501, Japan}








\begin{abstract}
Analyzing the X-ray data of supernova remnants (SNRs) are among the most challenging task in the current X-ray astronomy because SNRs are both spatially extended and variable over time. We developed the strategy to track the time-series properties of all the parts constituting a diffuse structure by introducing the {\it free-form image registration} technique based on B-spline, and demonstrated the methodology using the {\it Chandra} data of Cassiopeia~A. We successfully extracted the spatial distribution map of the time variability of continuum luminosity. To our knowledge, this is the first comprehensive characterization of such a dynamic diffuse target both in spatial and temporal viewpoints. We found that each of the four clusters derived by applying k-means algorithm to the extracted light curves has a clear physical meaning distinct from other clusters, which shows that our method is not a mere technique for automation but capable of capturing the underlying physics.
\end{abstract}



\section{Introduction} \label{sec:intro}
Through X-ray observations, several physical quantities can be obtained, such as positions, lightcurves, energy spectra, and polarizations. In many cases, however, only a low-dimensional slice of the complete dataset is essential. For example, the data of celestial point sources are essentially two-dimensional; one only needs to play with the spectra and lightcurves (energy $E$ and time $t$) because the spatial dimensions can be ignored --- they have no observable spatial substructures by definition, and their locations are usually unchanged. Effectively three-dimensional data (two spatial dimensions $(x,y)$ and energy $E$) are obtained by the observations of the objects that are spatially extended but stable in human timescale such as galaxy clusters. As supernova remnants (SNRs) are spatially extended and variable over human timescale, the data obtained through the X-ray observation of SNRs are truly four-dimensional ($x,y,t$ and $E$). In this regard, analyzing SNR data is among the most challenging task in the current X-ray astronomical data analyses.

When one wants to characterize the properties of a single diffuse system from a spatially comprehensive viewpoint (e.g., making the spatial distribution map of elemental abundances), typically a two-step strategy has been taken; (i) first, defining multiple regions so that they cover the entire system, and then (ii) performing the same analysis for all the regions. As both the steps can usually be automated, there have been many studies in this line \citep[e.g.,][]{ichinohe15,ichinohe17,ichinohe19,ichinohe21}.

On the other hand, when one wants to characterize the time-series properties of the specific structure in the object (temporally comprehensive viewpoint; e.g., extracting the time variability of the spectral indices of a hotspot), one needs to define the regions in all the time frames corresponding to that structure. This is not a difficult task when the object is stationary (e.g. distant point sources) because all the regions should be identical among observations. Even investigating the time-series properties of all the regions (comprehensive both spatially and temporally) in a stationary diffuse target (e.g. clusters of galaxies) is, at least conceptually, as easy as performing it for a single feature.

However, if the structures are moving or changing their shapes in the object, the analysis that is comprehensive in both spatial and temporal viewpoint (e.g., tracking the time variability of every part in a dynamic diffuse object) is, in turn, a very complex task for the following reasons. Firstly, even for a single distinct feature, defining the corresponding regions in all the time frames is usually not straightforward and requires human efforts to make them consistent and appropriate through the time frames. The situation is even worse when a region segmentation algorithm is employed to divide the field of view into multiple regions; as the algorithm is operated on each image independently, it is not guaranteed that the number of the regions, the topology of the regions (i.e., how the regions are connected to each other), and what fraction of the system each region represents, are same between any two images; Even if the algorithm guarantees all the things described in the previous sentence, associating the regions in two different images is nontrivial in itself.

Indeed, such time-series analyses have been performed only on the prominent features e.g. \citep[e.g.,][]{2007Natur.449..576U,2008ApJ...677L.105U,2009ApJ...697..535P,2014ApJ...789..138P,2020PASJ...72...85M,2022ApJ...940..105M}. This indicates that we might have missed important things happening in the regions that have not been analyzed yet. In order not to miss intriguing phenomena in the available data, the method that is able to capture the properties of the entire system in an unbiased manner is required.

To improve this situation and not to leave significant amount of data unexplored, we have developed the strategy to track the time-series properties of all the parts constituting a dynamic diffuse structure. The key idea is to find the segmentation of a given image that is same in the physical sense as a given segmentation of the reference image. We implement the strategy and demonstrate it using the multiple {\it Chandra} images of the supernova remnant Cassiopeia~A (hereafter Cas~A). 

X-ray emissions of Cas~A are known to be very complicated, where a mixture of thermal and non-thermal emissions and their temporal variations are observed in the monitoring observations since the launch of {\it Chandra} in 1999 \citep[e.g.,][]{2000ApJ...528L.109H,2011ApJ...729L..28P,2014ApJ...789..138P,2012ApJ...746..130H,2017ApJ...836..225S}. X-ray emitting structures in the remnant show different time variations while changing its position, making it difficult to know what causes the time variability in each component. Therefore, this remnant would be the best target for demonstrating our new technique. The outline of this paper is as follows. In Section~\ref{sec:char}, we explain the concept of our new strategy to track the time-series properties of the entire diffuse system. In Section~\ref{sec:demo}, we present how we implement the strategy, and demonstrate its effectiveness by applying it to the {\it Chandra} data of Cas~A. We discuss the results in Section~\ref{sec:discussion} and present the conclusion in Section~\ref{sec:conclusions}.

\section{Spatio-temporal characterization}\label{sec:char}

The main part of the algorithm consists of two steps; (i) finding the transformation that morphs a given image so that it matches the reference image best, based on the {\it B-spline registration} algorithm \citep{lee96,lee97}; (ii) using the transformation, inverse-transforming the image segmentation in the reference image's coordinates into the corresponding one in the given image's coordinates.

\subsection{Free-form image registration using B-spline}\label{sec:deform}
{\it Image registration} is the task of finding the transformation that maps any point in an image to the corresponding point in another image. Image registration methods are well studied in the medical field where comparing two images of the same object taken in different situations is essential \citep[e.g., two MRI images taken before and after the injection of a contrast agent; see e.g.,][]{rueckert99,mattes03}.

Rigid transformation is the simplest option. When the image is two-dimensional, the transformation only has three degrees of freedom corresponding to a rotation and two translations. Affine transformation is more general and has extra three degrees of freedom that describe the scaling and shearing. Although these methods are simple and easy to implement, they have the critical shortcoming that they can capture only the global motion.

It is often the case that the substructures in an astrophysical object changes non-uniformly across the field-of-view; for example, in an SNR, nonthermal filamentary shells tend to move outward from the center of explosion, while inner thermal substructures sometimes move inward. Moreover, often the velocities of the substructures are different due to the difference in the actual three-dimensional distance from the center as well as the surrounding environments.

In such a complex case, simple rigid transformation is severely insufficient and a more flexible transformation method is necessary. The main requirements to the method are; (1) it should be able to capture the local non-uniformity of the motions across the field of view, and at the same time, that (2) it can express the smoothness of the motion field. The latter point is important because although each substructure moves rather independently, the motion of a pixel would be similar to its neighboring pixels as astrophysical objects such as SNRs evolve in time mostly continuously.

There have been several methods to realize such {\it free-form image registration}. For example, the combination of feature detection \citep[e.g.,][]{lowe99,rublee11} and feature matching is often used to find corresponding key points in the two given images. However, X-ray images are usually dominated by the Poisson noises and image binning or smoothing is necessary to avoid detecting image fluctuations as feature points. This worsens the image resolution and thus is a significant drawback especially when the image is high-resolution such as the one taken with {\it Chandra}.

Another option is optical flow \citep[e.g.,][]{lucas81}, the algorithm developed for motion tracking in computer vision. However, optical flow is designed to take into account only the local information, namely, each pixel is independently assumed to be moving to a certain direction. Therefore, although this can be a choice when the motion of a local independent structure is focused on \citep[e.g.,][]{2018ApJ...853...46S}, this is not the best when the global consistency of the whole system such as smoothness of the motion field should be considered.

B-spline registration \citep{lee96,lee97} is an image registration algorithm that has been applied to the motion analysis of medical images. The basic concept of the method is to express the deformation field of the image using a finite number of discrete control points. The control points are arranged in a latticed pattern and each one represents the local motion of its neighboring coordinates. The motion field of the points between the control points are interpolated using cubic B-spline functions;

\begin{equation}
T_\mathrm{local}(x,y)=\displaystyle\sum^{3}_{k=0}\displaystyle\sum^{3}_{l=0}B_k(s)B_l(t)\phi_{i+k,j+l},
\end{equation}
where $i=\lfloor x\rfloor-1$, $j=\lfloor y\rfloor-1$, $s=x-\lfloor x\rfloor$, $t=y-\lfloor y\rfloor$ and $B_k$ represents the $k$th basis function of the B-spline
\begin{eqnarray*}
B_0(t)&=&(1-t)^3/6\\
B_1(t)&=&(3t^3-6t^2+4)/6\\
B_2(t)&=&(-3t^3+3t^2+3t+1)/6\\
B_3(t)&=&t^3/6,
\end{eqnarray*}
where $0\leq t < 1$. As the B-Spline representation of the motion field is continuous, differentiable, and bijective, this method is suitable for feature tracking of smoothly moving objects such as supernova remnants.

We want to identify the best transformation within this modeling. However, a caveat is that the definition of the {\it best} transformation is somewhat vague. It is straightforward to determine the best one when the ground truth deformation exists; that is, the best transformation should be the one that approximates the ground truth best within its expressive power. However, in the actual case, there is no ground truth deformation that transforms one observed image frame into another. Instead, there are just motions of astrophysical structures -- some move coherently, others independently --, and we want to {\it express} these collective motions by a parameterized transformation.

Therefore, in this work, we {\it define} the best transformation as the one that yields the most similar transformed image to the reference image, i.e., the one that minimizes a certain similarity metric predetermined by the analyzer of the data. In order to find the best transformation, one only needs to minimize the similarity metric, which measures the difference between the reference image and the other image after deformation. As this transformation is parameterized with a relatively small number of free parameters, i.e., the motion vector associated with each of the control points, it is relatively easy to find the best parameters with common optimization algorithms.

In some problematic cases, it is possible that the algorithm associates points in different images that are independent\footnote{For example, when three points are arranged in a triangle and all the points are rotating clockwise at the angular speed of 119$^\circ$ per frame, the optimizer is likely to halt by finding 1$^\circ$ per frame anticlockwise motion because the optimization step starts from zero motion.}. It should be noted that this method implicitly assumes the smallest possible motions. We expect that this assumption applies to most astrophysical applications, including the present work.

\subsection{Region transformation}
Once the best transformation is found, the reference image and the other image after deformation are similar to each other. It is thus expected that the region segmentation in the reference image generated by a physical motivation can also be used as a physically-motivated region segmentation of the other image after deformation. Therefore, the regions in the other image's original coordinates can be obtained by simply inverse-transforming the regions in the reference image. The resulting region segmentation should be physically motivated as is the one in the reference image.

\section{Demonstration} \label{sec:demo}
\subsection{Datasets} \label{sec:data}
\begin{table}[]
    \centering
    \begin{tabular}{ccrc}
        Year &  ObsID & Exp. (ks)   & Obs. Date    \\\hline
        2000 &  114   & 49.9   & 2000 Jan 30  \\\hline
        2002 &  1952  & 49.6   & 2002 Feb 06  \\\hline
        2004 &  4634  & 148.6  & 2004 Apr 28  \\
             &  4635  & 135.0  & 2004 May 01  \\
             &  4636  & 143.5  & 2004 Apr 20  \\
             &  4637  & 163.5  & 2004 Apr 22  \\
             &  4638  & 164.5  & 2004 Apr 14  \\
             &  4639  & 79.0   & 2004 Apr 25  \\
             &  5196  & 49.5   & 2004 Feb 08  \\
             &  5319  & 42.3   & 2004 Apr 18  \\
             &  5320  & 54.4   & 2004 May 05  \\\hline
        2007 &  9117  & 24.8   & 2007 Dec 05  \\
             &  9773  & 24.8   & 2007 Dec 08  \\\hline
        2009 &  10935 & 23.3   & 2009 Nov 02  \\
             &  12020 & 22.4   & 2009 Nov 03  \\\hline
        2010 &  10936 & 32.2   & 2010 Oct 31  \\
             &  13177 & 17.2   & 2010 Nov 02  \\\hline
        2012 &  14229 & 49.1   & 2012 May 15  \\\hline
        2013 &  14480 & 48.8   & 2013 May 20  \\\hline
        2014 &  14481 & 48.4   & 2014 May 12  \\\hline
        2015 &  14482 & 49.4   & 2015 Apr 30  \\\hline
        2016 &  18344 & 25.8   & 2016 Oct 21  \\
             &  19903 & 24.7   & 2016 Oct 20  \\\hline
        2017 &  19604 & 49.5   & 2017 May 16  \\\hline
        2018 &  19605 & 49.4   & 2018 May 15  \\\hline
        2019 &  19606 & 49.4   & 2019 May 13  \\\hline
    \end{tabular}
    \caption{{\it Chandra} observations}
    \label{tab:obsids}
\end{table}

The Advanced CCD Imaging Spectrometer (ACIS) of Chandra has observed Cas~A multiple times since its launch in 1999 \citep[e.g.,][]{2000ApJ...528L.109H,2000ApJ...537L.119H,2004ApJ...615L.117H,2008ApJ...677L.105U,2011ApJ...729L..28P,2014ApJ...789..138P,2017ApJ...836..225S,2018ApJ...853...46S} and here we used the archival data from 2000 to 2019. The ObsIDs used for this work are summarized in Table~\ref{tab:obsids}.  We reprocessed the archival level 1 event lists produced by the {\it Chandra} pipeline in the standard manner using the \verb+CIAO+ software package (version 4.12) and the \verb+CALDB+ (version 4.9.2.1) to apply the appropriate gain maps and the latest calibration products.

Cas~A is observed as a complex mixture of thermal and non-thermal X-ray radiations. For simplicity, we generated one image in the 4.2--6.0\,keV band for each observation, resulting in fourteen images in total. In this energy band, both thermal bremsstrahlung and non-thermal (synchrotron) radiation are present, and these featureless continuum radiations are dominant. Avoiding emission lines from various elements reduces information on specific elements (i.e., information on the elemental abundance). This allows us to focus only on thermal and non-thermal variations.

\subsection{Implementation}\label{sec:impl}
We implemented the cubic B-spline registration using \verb+SimpleITK+ toolkit \citep{lowekamp13,beare18,yaniv18}. We employed an 8$\times$8 mesh for the control points in the transform domain. Each control point has two control parameters corresponding to the two components of the motion field vector there in the two-dimensional image, and the cubic B-spline interpolation requires three extra control points per dimension for the interpolation of the domain close to the image boundaries. These result in $(8+3)\times(8+3)\times2=242$ parameters to optimize. We employed the L-BFGS-B algorithm implemented in \verb+SimpleITK+ for the metric optimization \citep{liu89,byrd95,zhu97}. For the optimization metric, the correlation of two images is used because the luminosity of Cas~A is known to be gradually decreasing \citep[e.g.,][]{2011ApJ...729L..28P}, and simpler, intensity-based metrics such as mean-squared error are not suitable.

In this work, we use the image obtained in the 2004 observation as the reference image because it is of the highest quality. We preprocessed all the fourteen images with a Gaussian filter of the kernel size $\sigma=4$~px prior to the deformation. For each of thirteen images other than the reference, the corresponding transformation was derived by comparing it with the reference image. 

Although all the images are directly compared to the 2004 image for deriving the transformations (one-shot strategy) in the present implementation, it should be noted that another strategy exists; the two-shot strategy, in which the transformation between every two successive frames is first computed, and the appropriate transformations are composed to reproduce the transformations equivalent to the one-shot counterparts. We checked both strategies and found that both yield almost the same results, while some of the structures show more residual motions when the two-shot strategy is taken. We think this is due to the larger uncertainties in the determination of the transformation when using worse-quality images. There are at least two competing factors that cause larger errors in the resulting transformations; (1) larger time intervals and (2) much accumulation of the errors. In the present case, we think that the latter effect dominates so that the one-shot strategy works better. While the 2004 image has an exposure time of $\sim$1~Ms, other images have exposure times of only $\sim$50~ks (Table~\ref{tab:obsids}). In the two-shot strategy, most of the transformations are computed using these worse-quality images, which would result in larger errors for each transformation. In the present case, the motions of the structures are relatively smooth and slow, and we thus think the accumulation of the errors has a larger negative impact on the overall uncertainties compared to the larger time intervals. If, for example, the motions of the structures are fast, the time interval is very large, and/or the quality of all the images is very high (or low), the former factor could dominate. In such cases, the two-shot strategy would work better.

\begin{figure*}[ht!]
\plotone{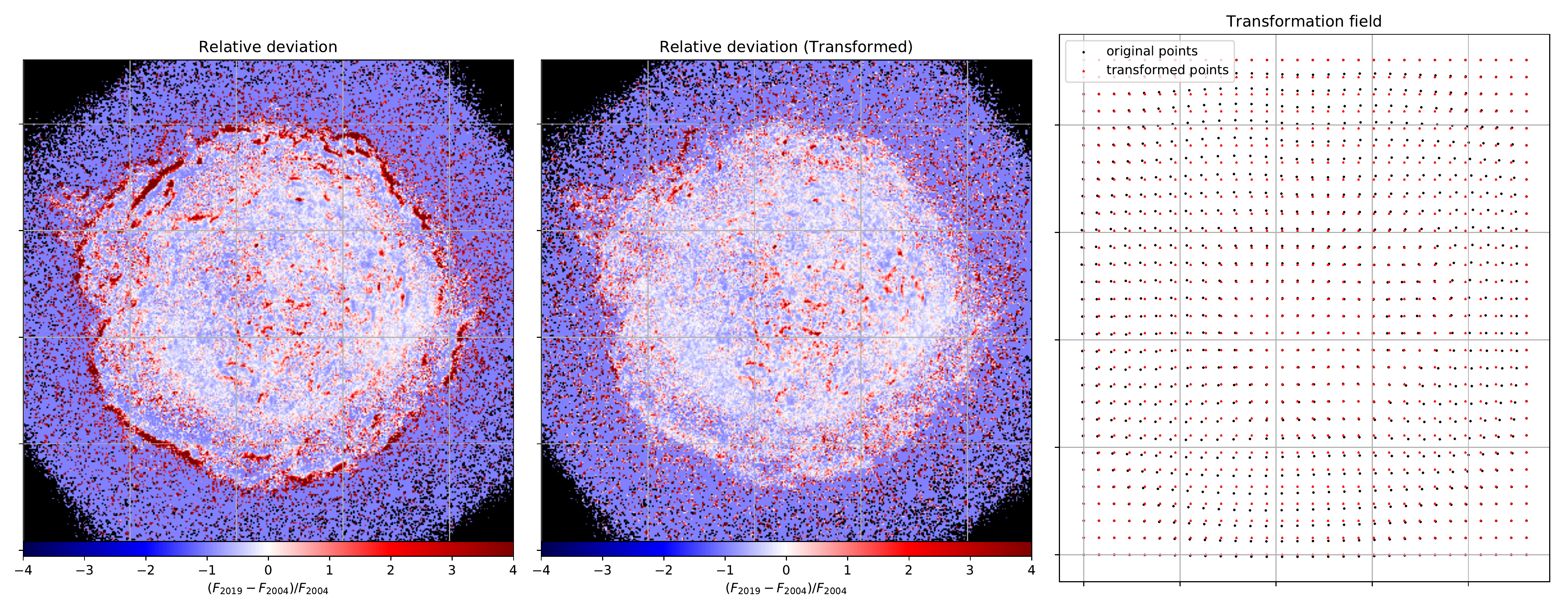}
\caption{{\it Left and middle:} relative deviation images computed by using $(X-Y)/Y$ with $X$ and $Y$ being the 2019 and 2004 (reference) images, respectively. {\it Left:} created using the original images (without any transformations). {\it Middle:} the 2019 image is transformed to match the 2004 image best. {\it Right:} visualization of the transformation field. The coordinates in the 2019 image represented by the black points are transformed to the coordinates represented by the red points. Note that, in the left and middle panels, most of the pixels outside the remnant show negative deviations because the 2004 image contains far fewer zero-count pixels due to the long exposure time ($\sim$1~Ms, see also Table~\ref{tab:obsids}).} \label{fig:comp_reldiv}
\end{figure*}

The left panel in Fig.~\ref{fig:comp_reldiv} shows the relative deviation image computed by using $(X-Y)/Y$, where $X$ and $Y$ is the images obtained in the 2019 and 2004 (reference) observations, respectively. The image shows clear filamentary structures aligned in a circular shape, reflecting the non-thermal filaments in Cas~A, which are actually moving outwards. The image in the middle panel is computed by the same formula, in which the 2019 image deformed using B-spline so as to match the 2004 image best is used as $X$, instead of the original 2019 image. The disappearance of the filaments in all directions clearly shows that the 2019 image is successfully deformed by the B-spline transformation. The right panel shows a visualization of the transformation field. The coordinates in the 2019 image represented by the black points are transformed to the ones represented by the red points. It can be observed that in addition to the dominant radial motions of the filaments, the non-uniformity of their amplitudes and directions are captured, demonstrating the effectiveness of this method on the registration problem of SNR images.

To divide the field of view into subregions, we used the contour binning algorithm \citep{sanders06}. We applied the algorithm to the reference image with the parameters \verb+sn=50+, \verb+constrainval=2.5+, and \verb+smoothsn=15+. The region segmentation in the reference image's coordinates was converted into the ones in all the other images' original coordinates by the respective inverse transformations.

Once the regions are obtained in all the images, one can easily obtain the time-series property associated to any features; one only needs to refer to the region expressed in the coordinates corresponding to the image of interest without being bothered by, e.g., the consistency of the regions among different time frames. For the demonstration purpose, we simply extracted the continuum X-ray flux in each region for all the time frames. As a result, $\sim$2000 light curves, each of which corresponds to a certain morphological structure in Cas~A, were obtained. Using the simulated image frames, we tested the validity of the pipeline and conservatively estimated the systematic uncertainties in deriving the light curve values at $\sim10\%$. See Appendix~\ref{sec:app1} for the details.

For the current implementation, most of the calculations take less than a few seconds using an Intel Xeon E5-1620 v4 CPU (3.50~GHz, 8-core). The processes that can take longer time are the image registration and the inverse transformation of the region map into the target coordinates. With the image dimensions of 921$\times$921, the typical time the former process takes ranges from about ten seconds to several tens of seconds per pair of images depending on the registration parameters. The latter takes $\sim$20~seconds per target frame.

\subsection{Unsupervised clustering}\label{sec:clustering}
To characterize the thousands of light curves, we adopted the k-means clustering algorithm \citep{macqueen67}. k-means is one of the unsupervised machine learning algorithms to perform the partition of $N$ observations contained in a given dataset into $k$ clusters. After successful classification, each element belongs to one of the $k$ clusters whose center is the nearest to the element among the $k$ centers. The center of each cluster is calculated by the mean of the elements belonging to the cluster, which are expected to be similar to each other. We ran k-means with $k=4$ with the Euclidean metric (see Appendix~\ref{sec:app2} for the details regarding the choice of $k$). Before applying k-means, each of the light curves was normalized using the respective 2004 value.

\begin{figure*}[ht!]
\plotone{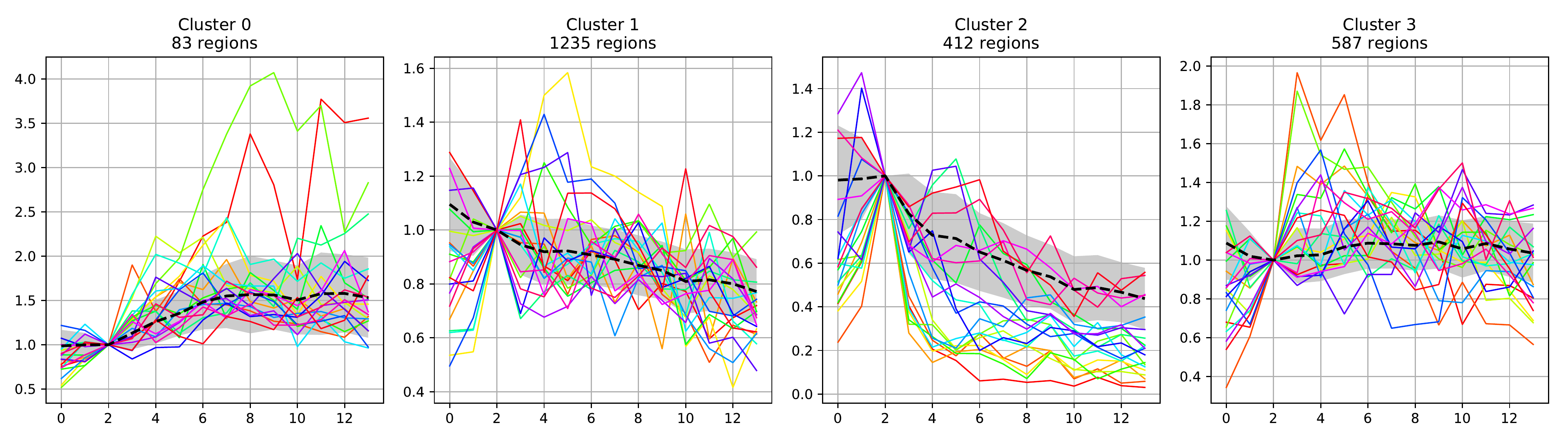}
\caption{The results of k-means ($k=4$). Each panel corresponds to a cluster. In each panel, the colored lines exemplify the light curves belonging to the cluster, and the black dashed line and the gray band correspond to the average light curve and standard deviation, respectively.\label{fig:lcs}}
\end{figure*}

Fig.~\ref{fig:lcs} shows the result of k-means. The average light curve (black dashed line) of each cluster represents the rough trend of the light curves belonging to the cluster. Namely, the X-ray flux of the regions belonging to the clusters 0, 1, 2, and 3 is increasing, mildly decreasing, quickly decreasing, and staying constant, respectively.

\subsection{Visualization}\label{sec:vis}

\begin{figure*}[ht!]
\includegraphics[angle=90,width=17cm]{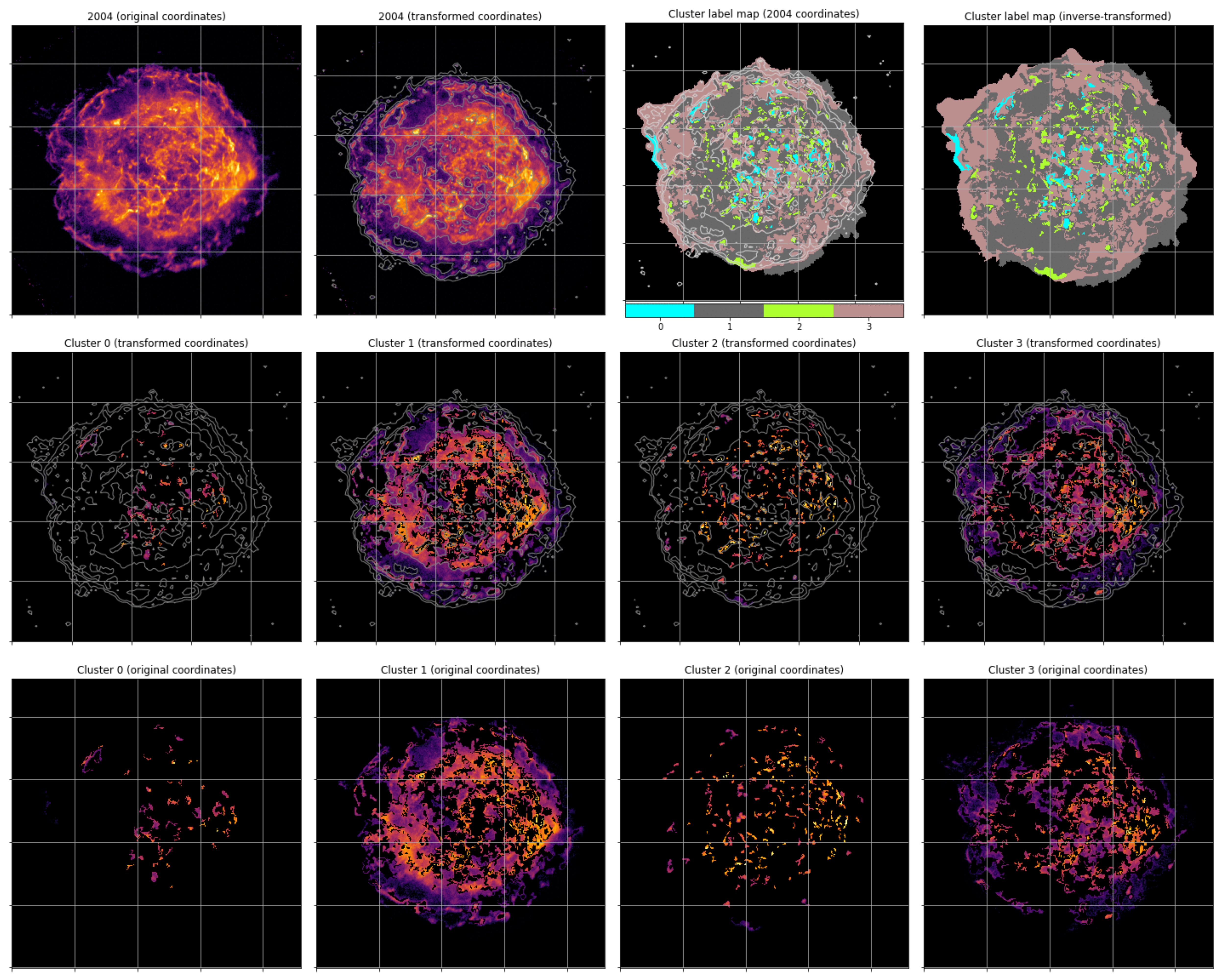}
\caption{Visualization of the results. {\it Top left:} the Cas~A X-ray images in the original coordinates. {\it Top, second from left:} the same images in the transformed coordinates. {\it Top, third from left:} the spatial distribution map of the clustering results in the 2004 coordinates. {\it Top, right:} the same maps inverse-transformed onto the target frames' coordinates. {\it Middle row:} the Cas~A X-ray images filtered by the clusters, shown in the transformed coordinates. The leftmost, second, third, and rightmost panels show the regions in Cas~A, whose light curves are classified as clusters 0, 1, 2, and 3, respectively. {\it Bottom row:} the same images as in the middle row, shown in the original coordinates. The contours represent the X-ray surface brightness at 2004. The animation (14 frames, corresponding to the years shown in Table~\ref{tab:obsids}) shows how the overall morphology of Cas~A evolves with time and how this morphological evolution is seemingly suppressed by our method.\label{fig:map}}
\end{figure*}

The resulting product of all the above procedures is a set of light curves, grouped by time variability. As each light curve represents a certain part in the object, one can convert this information into the spatial distribution map of X-ray time variability. The top right panel of Fig.~\ref{fig:map} shows the spatial distribution map of the clustering results. The middle and bottom rows of Fig.~\ref{fig:map} show the Cas~A images filtered by the cluster memberships. The components whose flux are increasing (cluster 0; middle left) or rapidly decreasing (cluster 2; bottom left) are distributed in clumpy morphology. On the other hand, the components whose flux are gradually decreasing (cluster 1; middle right) or constant (cluster 3; bottom right) are distributed in diffuse morphology.

Although both clumpy, the distribution of the brightening (cluster 0) and rapidly dimming (cluster 2) components are different; the former is rather localized in the northwestern direction, while the latter is rather distributed uniformly. The bright compact clumps tend to belong to the cluster 2, while the cluster 0 contains filamentary structures of moderate luminosity . Regarding the diffuse components, the gradually dimming component (cluster 1) is apparently brighter than the constant luminosity component (cluster 3). The regions covered by the cluster 1 extend to the radii of non-thermal shells, while the cluster 3 only covers the region inside the shells.

\section{Discussion} \label{sec:discussion}
\begin{figure*}[ht!]
\plotone{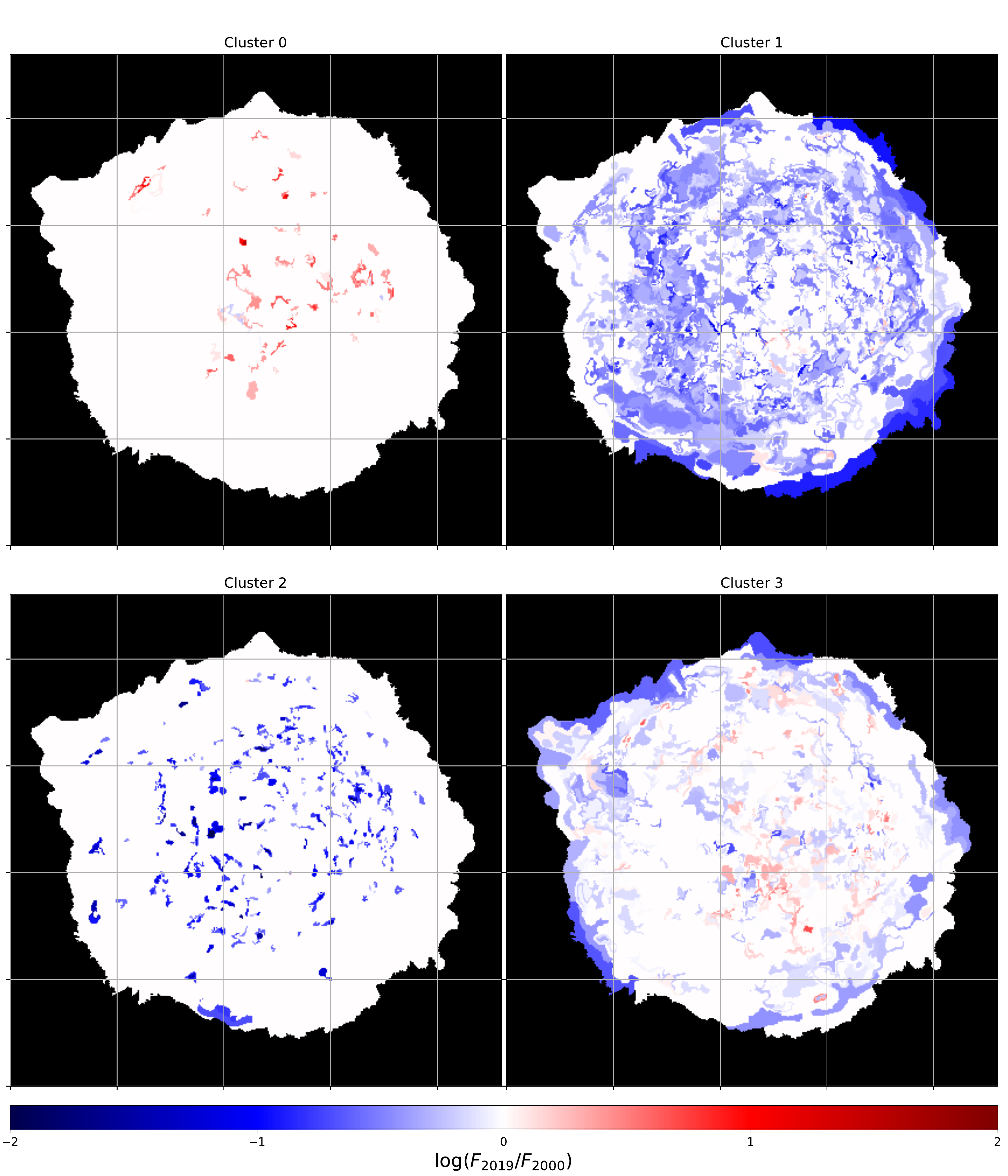}
\caption{The spatial distribution maps of the ratio of the flux in 2019 ($F_{2019}$) and 2000 ($F_{2000}$). In each panel, only the component classified as the respective cluster is shown. \label{fig:flmap}}
\end{figure*}
In the previous section, we have shown the concept of the combination of the B-spline registration and the k-means clustering algorithm in characterizing the evolving diffuse structures, and demonstrated the effectiveness of the concept using the data of Cas~A. To our knowledge, this is the first comprehensive characterization of such a dynamic diffuse target both in spatial and temporal viewpoints. In this section, we show that the method is not a mere tool for automation by demonstrating that the clusters thus obtained are actually scientifically interpretable. We also discuss the advantages, caveats and future prospects of our method.

\subsection{Physical interpretation of each cluster} \label{sec:interpretation}
We have identified four clusters showing different time variations. These can be broadly classified into two types of variations -- thermal or non-thermal emissions. Here, we discuss the origin of the flux variations using the typical timescale of each cluster.

Fig.~\ref{fig:flmap} show the spatial distribution maps of the ratio of the flux in 2019 and 2000, plotted for each of the clusters. We can see that the regions with flux increase and decrease between 2000 and 2019 were characterized well. Interestingly, the different rates of variation can also be visualized (e.g., between the clusters 1 (deep blue) and 2 (light blue)), suggesting that different physics is in the background.

Generally, non-thermal emissions in SNRs are observed as small-scale filamentary structures \citep[e.g.,][]{2003ApJ...589..827B,2005ApJ...621..793B}. This is because the non-thermal X-rays from the accelerated cosmic-ray electrons are emitted from small regions that are compressed by shock waves. Therefore, the small-scale knotty/filamentary structures seen in the clusters 0 and 2 would probably be associated with the non-thermal phenomena. In addition, these clusters show fast variability with $\sim +3.5$ \% yr$^{-1}$ ($e$-folding time $\tau_{\rm inc}\sim17$~yr) for the cluster 0 and $\gtrsim -4$ \% yr$^{-1}$ ($e$-folding time $\tau_{\rm dec}\sim49$~yr) for the cluster 2, which could be related to rapid acceleration and cooling in an amplified magnetic field \citep[e.g.,][]{2007Natur.449..576U,2008ApJ...677L.105U,2009ApJ...697..535P}. The synchrotron cooling timescale $\tau_{\rm syn}$ and the acceleration timescale $\tau_{\rm acc}$, which are $e$-folding timescales, can be estimated as
\begin{eqnarray}
\tau_{\rm syn} \,&\sim& \,0.7 \,B_{\rm mG}^{-1.5} \,\epsilon^{-0.5}_{5{\rm keV}} ~~{\rm year}\\
\tau_{\rm acc} \,&\sim& \,0.8 \,\eta \,B_{\rm mG}^{-1.5} \,\epsilon^{0.5}_{5{\rm keV}} V_{\rm sh,5}^{-2} ~~{\rm year},
\end{eqnarray}
where $B_{\rm mG}$, $V_{\rm sh,5}$, $\epsilon_{5{\rm keV}}$ and $\eta$ are the magnetic field in units of 1\,mG, the shock velocity in units of 5,000 km s$^{-1}$, the photon energy in units of 5\,keV, and the Bohm factor, respectively. We found that the amplified magnetic field of $B \sim$ 0.12\,mG and the Bohm factor of $\eta \sim$ 3 reproduce the observational timescales well, which agrees with the previous estimations \citep[e.g.,][]{2018ApJ...853...46S,2021ApJ...907..117T}. Thus, we conclude that most of the clusters 0 and 2 are related to the non-thermal phenomena.

We note that our method is also useful for identifying peculiar structures, such as inward-moving shocks in this remnant \citep{2018ApJ...853...46S}. These structures can be interpreted as signatures of the interaction of the remnant with an asymmetric dense circumstellar shell that occurred between $\sim$180 and $\sim$240 yr after the supernova event \citep{2022ApJ...929...57V,2022A&A...666A...2O}, thus it would be important for understanding the progenitor's activity. We found that these structures, classified as clusters 0 and 2 in the southern and western regions, move differently from most of the structures that move outward, but were successfully tracked by our method (see the rightmost panel of Fig.~\ref{fig:comp_reldiv} and the movie corresponding to Fig.~\ref{fig:map}). While there are also several bright small-scale structures classified as clusters 0 and 2 in the northwestern reverse-shock region, they differ from the inward-moving structures in that they are moving outward. This suggests that there is not enough material in the northwest direction to drive the shock wave inward, and may support the recently proposed asymmetric circumstellar shell for Cas A. In addition to these structures, our method identified a forward-shock filament that is getting brighter in the northeastern edge of the remnant (see the top left panel of Fig.~\ref{fig:flmap}). It has been suggested that diffusion in this filament appears to be occurring near the Bohm limit \citep{2006NatPh...2..614S}, which implies efficient cosmic-ray acceleration at the location. Although the interpretation of the cause of the flux variation is outside the scope of this paper, our analysis indicates that the flux increase in this filament reported in \cite{2009ApJ...697..535P} has continued since then.

Diffuse components that are gradually dimming with a rate of $\sim -1.4$ \% yr$^{-1}$ were classified into the cluster 1, which could be related to the thermal emission of Cas~A. In particular, the regions classified into this cluster appear to be concentrated in the northern and southeastern reverse shock regions where the thermal emission is dominant \citep[e.g.,][]{2004ApJ...613..343D,2008ApJ...686.1094H}. Assuming that this cluster represents the thermal radiation, we here discuss the time evolution of the thermal component in the remnant. The thermal emissions in young supernova remnants are thought to decrease due to adiabatic expansion. In \cite{2017ApJ...836..225S}, the decay rate of the thermal X-rays in 4.2--6.0 keV by the adiabatic expansion in Cas~A was estimated to be $-1.36~(m/0.66) (t/340~{\rm yr})^{-1}$ \% yr$^{-1}$, where $m$ and $t$ is the expansion index and the age of the remnant, respectively. This rate explains well the temporal variation of the cluster 1, thus we conclude that this cluster would be related to the thermal emission. On the other hand, we note that the non-thermal emissions dimming with a longer timescale (i.e., cooling at a less amplified magnetic field) would also be classified into this cluster.

Faint regions with less time variability were classified into the cluster 3. We consider that acceleration (or heating) and cooling are balanced in these regions. Most of the regions in this cluster are located at the southwestern reverse shock regions and the forward shock regions, where the non-thermal emission is dominant \citep[e.g.,][]{2004ApJ...613..343D,2008ApJ...686.1094H,2015ApJ...802...15G}. If cosmic-ray electrons were regularly accelerated and cooled there, this would explain the constant luminosity. In addition, materials are constantly being heated in the vicinity of the shock wave, which may offset the decay of the thermal component due to adiabatic expansion. These balanced components could have been classified as the cluster 3.

\subsection{Advantages, caveats and future prospects}
As shown in the previous section, we found that each of the cluster represents a clear physical meaning distinct from other clusters. This means that the physical motivation in each cluster is aligned and thus that our methodology can extract the physics underlying a dynamic diffuse object in an organized manner. Notably, all the processes, i.e., the region segmentation, image registration, and clustering, were performed automatically. This is already a huge advantage that lead to the first comprehensive characterization of Cas~A, which have been practically impossible to do manually. Such an automated extraction of physics would become more important when the future powerful missions such as {\it Athena} are in operation.

Most of the future missions, including {\it Athena}, will have worse angular resolutions than {\it Chandra}. Although worse angular resolutions cause worse image quality, we think the registration should still work if the object appears to change its shape under that quality. If the angular resolution is so poor that actually moving structures do not appear to be moving, our method is just unnecessary, and one can simply take identical sets of the regions for all the time frames.
In general, when the angular resolution is poor, one needs to be careful in determining the regions to take the PSF (point spread function) blending effect into consideration properly, and it is probably required to tune the hyperparameters of the tessellation algorithm. That said, we emphasize that this is the issue in the tessellation algorithm but not in the registration algorithm presented in this paper.

The concept of our method is fairly simple --- using image registration technique to convert dynamic objects into static ones. Therefore, at least conceptually, our method can be applied to a wide range of other dynamic diffuse targets, regardless of the instrument (e.g., {\it Chandra} or {\it XMM-Newton}; it does not even have to be X-ray instruments). That said, there are cases that our method would not work properly. For example, when the neighboring image frames are not very similar, image registration would fail. This happens, for example, when the structures are moving too fast compared to the time separation of two images or when the motion is chaotic. Low quality of images (low statistical counts, high noise, or existence of instrumental artefacts) would of course cause the fail of the method. When the control points are too sparse compared to the spatial scale of the motion, this method might also fail. The cases include the motions whose scale is substantially sharper than the mesh grid, and the motions which are highly non-uniform within the neighrboring of a single control point. This behavior is also observed in our result; some filamentary structures close to the center show residual motion after transformation (see the movie corresponding to Fig.~\ref{fig:map}). Although employing denser control points should resolve the problem, this leads to the increase of the number of parameters to optimize. Therefore careful trade-off study may be required.

In this demonstration, we used only the total luminosity of the continuum (4.2--6.0\,keV) band for both image transformation and the subsequent clustering. Although we leave advanced analysis for future works, we note that it would be easy to extend this method to incorporate more information contained in the dataset. For example, by extracting lightcurves from multiple energy bands, it would be possible to characterize the time evolution of the object from the time evolution of spatial distribution of metals. By using high-resolution spectroscopic data using future instruments, one can trace the evolution of line-of-sight velocity of moving structures to constrain 3D shock dynamics. Ultimate goal would be using the entire spectral information. Characterizing a huge amount of complex spectra is itself a significant task and several progresses have been made \citep[e.g.,][]{iwasaki19}. We think that combining our method with such methods based on data science would facilitate the analysis of future high-quality astronomical datasets.

\section{Conclusions} \label{sec:conclusions}
We have developed the strategy to track the time-series properties of all the parts constituting a diffuse structure by introducing the free-form image registration technique based on B-spline. We have demonstrated the methodology using the {\it Chandra} data of Cas~A by extracting the spatial distribution map of the time variability of continuum luminosity. To our knowledge, this is the first comprehensive characterization of such a dynamic diffuse target both in spatial and temporal viewpoints.

We have found that each of the four clusters derived by applying k-means algorithm to the extracted light curves has a clear physical meaning distinct from other clusters, which shows that our method is not a mere technique for automation but capable of capturing the underlying physics.

\begin{acknowledgments}
This work was partially supported by the Grants-in-Aid for Scientific Research by the Japan Society for the Promotion of Science with KAKENHI Grant Nos. 18H05458, 19K14749, 20K14524, and 20K20527.
\end{acknowledgments}

\appendix
\section{Test of the pipeline}\label{sec:app1}
In this section, we present the results of the experiment to test the validity of the presented algorithm and to estimate the systematic uncertainties in the light curve reconstruction.

\begin{figure*}[ht!]
\plotone{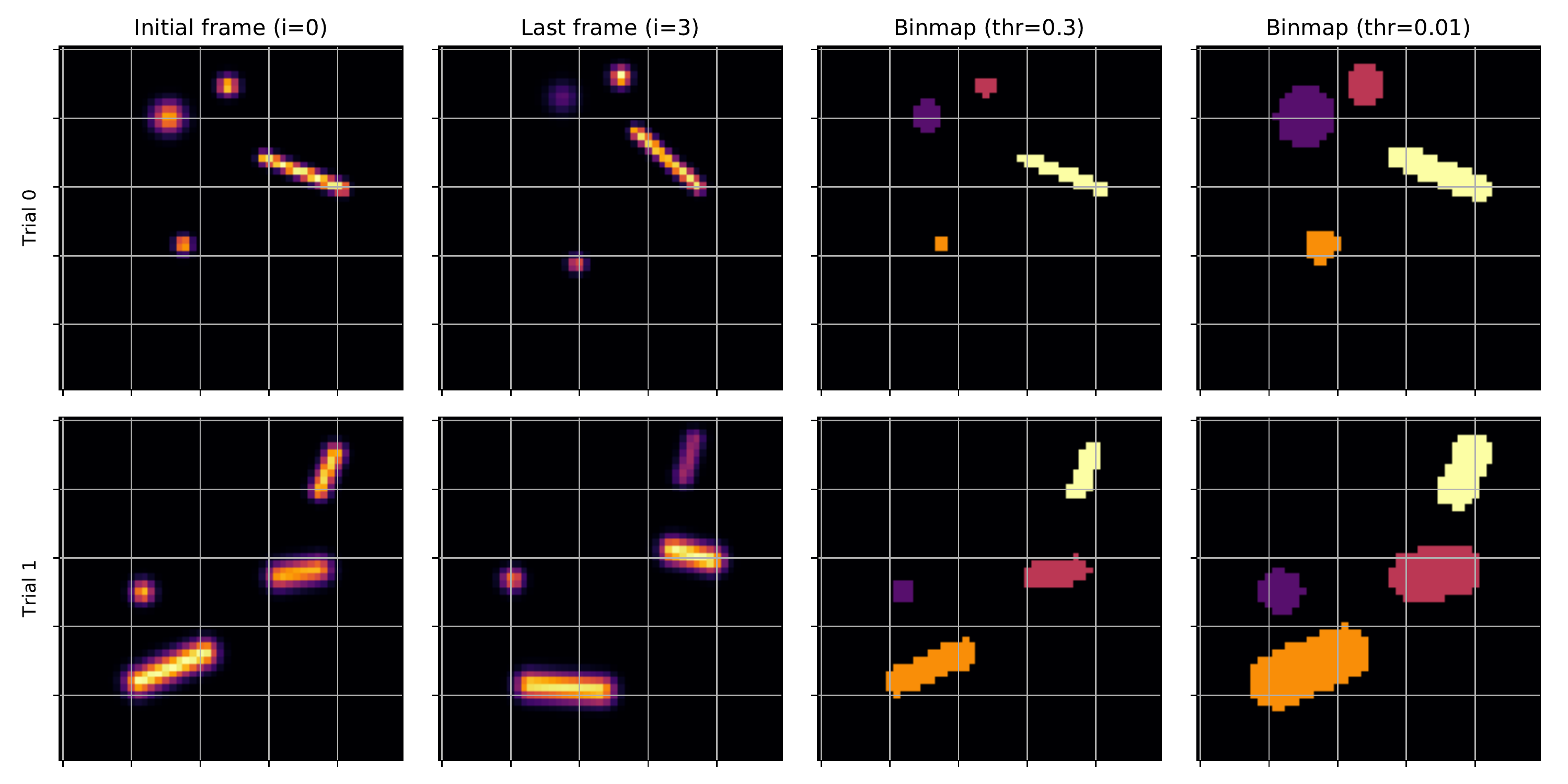}
\caption{Examples of the first and last frames of randomly generated mock movies (left two columns) and the corresponding region maps with the threshold values of 0.3 and 0.01 (right two columns). The first frames are presented in the left column, and the corresponding last frames in the second. {\it Top row:} three clumps and a filament. {\it Bottom row:} a clump and three filaments.\label{fig:app1}}
\end{figure*}

We simulated 50 movies consisting of four 50~px$\times$50~px mock image frames and randomly generated four structures per movie. Each structure is either a `clump' or a `filament'. For a given frame number $i$ ($i=0,1,2,3$), the shape of a clump is computed by a two-dimensional Gaussian with the center location, normalization, and radius calculated by ($x_0+i\Delta x$,$y_0+i\Delta y$), $n_0+i\Delta n$, and $\sigma$, respectively, using the values presented in Table~\ref{tab:app1}. The shape of a filament is a line segment smeared by a two-dimensional Gaussian. The center location, rotation, normalization, Gaussian radius, and line length are calculated by ($x_0+i\Delta x$,$y_0+i\Delta y$), $\theta_0+i\Delta\theta$ $n_0+i\Delta n$, $\sigma$, and $l$, respectively, using the values presented in Table~\ref{tab:app1}. The left two columns of Fig.~\ref{fig:app1} show examples of the first frames of the movies and the corresponding last frames.

Using the pixel values corresponding to each structure, we defined three regions for each structure by selecting the pixels whose value is larger than 0.3, 0.1, and 0.01 times the maximum. Larger threshold values correspond to smaller region sizes, as exemplified in the right two columns of Fig.~\ref{fig:app1}. We found that sometimes those randomly generated structures overlap, which results in the regions not being properly determined. To avoid such situations, we only kept the random trials that did not cause the overlap of regions.

\begin{figure}[ht!]
\plotone{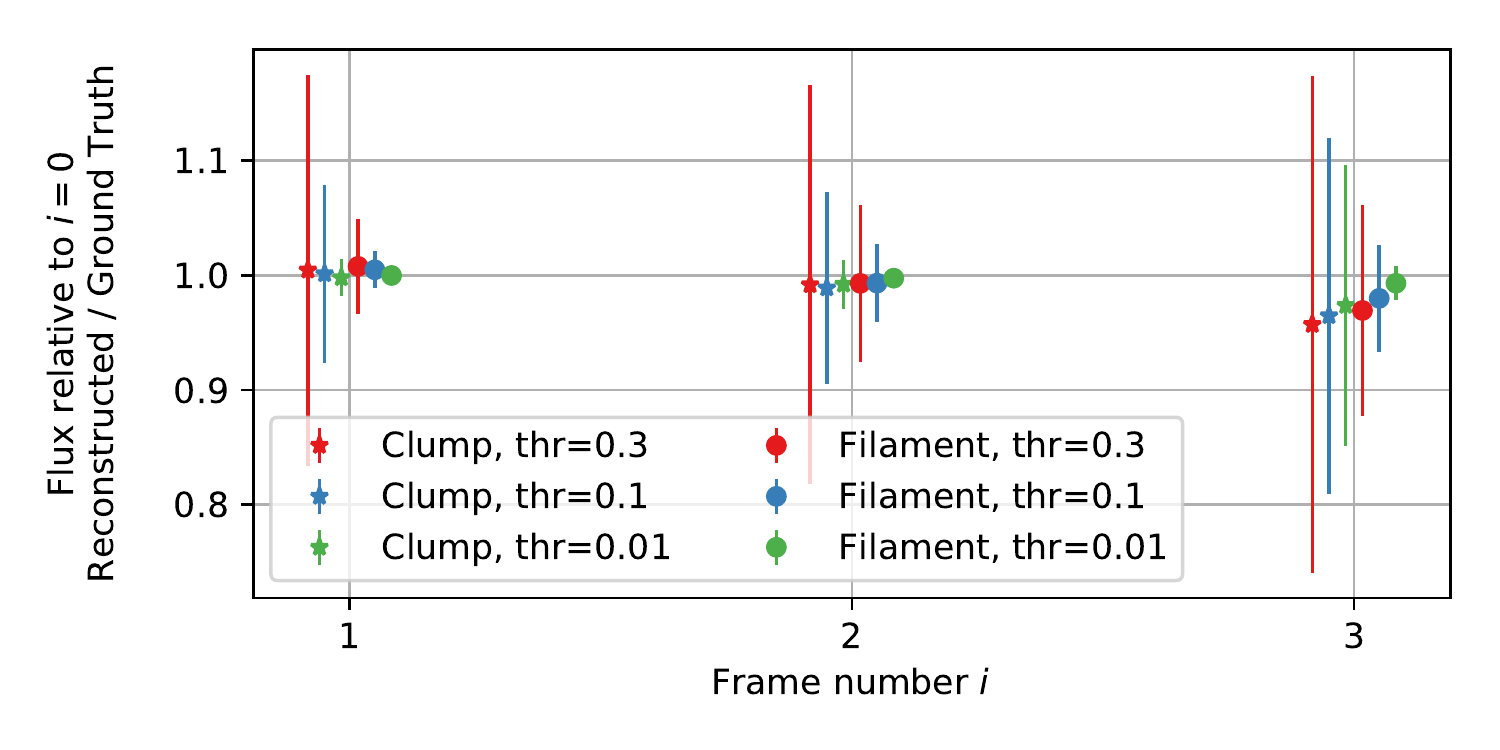}
\caption{$\hat{n}_{i,\mathrm{recon}}/\hat{n}_{i,\mathrm{GT}}$, the ratio of the reconstructed normalization value ($\hat{n}_{i,\mathrm{recon}}$) to the ground truth normalization value ($\hat{n}_{i,\mathrm{GT}}$). Before calculating the ratio, all the values are normalized by the corresponding normalization value at the first frame (i.e., $\hat{n}_i\equiv n_i/n_0$). Different colors encode different region thresholds, and different markers different structure shapes.\label{fig:app2}}
\end{figure}

Using the region maps, we followed the same procedure presented in Section~\ref{sec:char} to transform the region maps into the coordinates of target frames and extracted the sum of the pixel values corresponding to the transformed regions. Fig.~\ref{fig:app2} shows the ratios of the reconstructed normalization values to the ground truth normalization values. The errors correspond to the standard deviation of the derived values.

The ratios are consistent with unity, suggesting that the light curves are recovered correctly by our strategy. We found that the scatter of the reconstructed values increases with frame numbers, which can be naturally interpreted as that it is easier to perform image registration between two similar frames than two rather different frames. We also found that the scatters are larger for higher threshold values (i.e., smaller regions). This is because the errors in the deformation cause the misalignment of structures and the corresponding regions. The systematic errors introduced by this effect should be larger for smaller regions because the relative error of the transformation field to the region size becomes larger.

It is difficult to quantitatively estimate how much error is caused by how the structures in different frames differ because other factors, such as the distance and flux ratio to the surrounding structures, would also affect the predictions. However, considering that our simulation includes the structures whose motion scales are as large as $\sim1/10$ of the frame size (e.g., the filaments in Fig.~\ref{fig:app1}), which is way larger than the structure motions existing in Cas~A, we think the overall systematic errors due to the reconstruction accuracy is sufficiently below the error bars presented in Fig.~\ref{fig:app2}, i.e., a conservative baseline value is $\sim10\%$. We expect that such a level of error does not significantly affect the k-means algorithm (which resulted in four clusters of rather distinct trends, see Fig~\ref{fig:lcs}) and the following discussions.

\begin{table*}[]
    \centering
    \begin{tabular}{lrr}
        Parameter & Distribution (clump) & Distribution (filament) \\\hline
        $x_0$, $y_0$ & Uniform(5, 45) & Uniform(7.5, 42.4) \\
        $\Delta x$, $\Delta y$ (frame$^{-1}$) & Uniform(0.3, 1.0) & Uniform(0.3, 1.0)\\
        $\theta_0$ (deg) & N/A & Uniform(0, 360) \\
        $\Delta\theta$ (deg~frame$^{-1}$) & N/A & Uniform(-10, 10) \\
        $n_0$ & Uniform(1.0, 1.5) & Uniform(1.0, 1.5) \\
        $\Delta n$ (frame$^{-1}$) & Uniform(-$n_0$/6, $n_0$) & Uniform(-$n_0$/6, $n_0$)\\
        $\sigma$ & Uniform (0.5, 1.5) & Uniform (0.5, 1.5)\\
        $l$ & N/A & Uniform(5, 15) \\
    \end{tabular}
    \caption{Parameter distribution to generate random structures.}
    \label{tab:app1}
\end{table*}

\section{Note on the k-means hyperparameters}\label{sec:app2}
In this work, we ran k-means with $k=4$ with the Euclidean metric.
In determining the number of clusters, we checked the performance metrics of the k-means algorithm, such as the silhouette score \citep{ROUSSEEUW198753}, the elbow method, and the Bayesian information criterion \citep[BIC,][]{10.1214/aos/1176344136,10.5555/645529.657808}. Although all the methods suggested the optimal number of around $k=3-5$, we found it difficult to choose one. We think this is because the light curve properties are relatively smoothly distributed, and there are no clear boundaries between the components.
Therefore, we arbitrarily chose $k=4$ because we found that $k=5$ yields a cluster containing only $\sim$30 samples, and that $k=3$ failed to separate the brightening component from the constant luminosity component.

For a situation like this, k-means is probably not the optimal algorithm, and other algorithms that are capable of characterizing such a smoothly distributed dataset would be more suitable. Exploring the k-means hyperparameters (which include not only the number of clusters, but also e.g., the distance measure and how to calculate the center) and/or the optimal algorithm is beyond the scope of this paper.

\bibliography{ref}{}

\begin{thebibliography}{}
\expandafter\ifx\csname natexlab\endcsname\relax\def\natexlab#1{#1}\fi
\providecommand{\url}[1]{\href{#1}{#1}}
\providecommand{\dodoi}[1]{doi:~\href{http://doi.org/#1}{\nolinkurl{#1}}}
\providecommand{\doeprint}[1]{\href{http://ascl.net/#1}{\nolinkurl{http://ascl.net/#1}}}
\providecommand{\doarXiv}[1]{\href{https://arxiv.org/abs/#1}{\nolinkurl{https://arxiv.org/abs/#1}}}

\bibitem[{{Bamba} {et~al.}(2003){Bamba}, {Yamazaki}, {Ueno}, \&
  {Koyama}}]{2003ApJ...589..827B}
{Bamba}, A., {Yamazaki}, R., {Ueno}, M., \& {Koyama}, K. 2003, \apj, 589, 827,
  \dodoi{10.1086/374687}

\bibitem[{{Bamba} {et~al.}(2005){Bamba}, {Yamazaki}, {Yoshida}, {Terasawa}, \&
  {Koyama}}]{2005ApJ...621..793B}
{Bamba}, A., {Yamazaki}, R., {Yoshida}, T., {Terasawa}, T., \& {Koyama}, K.
  2005, \apj, 621, 793, \dodoi{10.1086/427620}

\bibitem[{Beare {et~al.}(2018)Beare, Lowekamp, \& Yaniv}]{beare18}
Beare, R., Lowekamp, B., \& Yaniv, Z. 2018, Journal of statistical software,
  86, 8.
\newblock \url{https://pubmed.ncbi.nlm.nih.gov/30288153}

\bibitem[{Byrd {et~al.}(1995)Byrd, Lu, Nocedal, \& Zhu}]{byrd95}
Byrd, R.~H., Lu, P., Nocedal, J., \& Zhu, C. 1995, SIAM J. Sci. Comput., 16,
  1190

\bibitem[{{DeLaney} {et~al.}(2004){DeLaney}, {Rudnick}, {Fesen}, {Jones},
  {Petre}, \& {Morse}}]{2004ApJ...613..343D}
{DeLaney}, T., {Rudnick}, L., {Fesen}, R.~A., {et~al.} 2004, \apj, 613, 343,
  \dodoi{10.1086/422906}

\bibitem[{{Grefenstette} {et~al.}(2015){Grefenstette}, {Reynolds}, {Harrison},
  {Humensky}, {Boggs}, {Fryer}, {DeLaney}, {Madsen}, {Miyasaka}, {Wik},
  {Zoglauer}, {Forster}, {Kitaguchi}, {Lopez}, {Nynka}, {Christensen}, {Craig},
  {Hailey}, {Stern}, \& {Zhang}}]{2015ApJ...802...15G}
{Grefenstette}, B.~W., {Reynolds}, S.~P., {Harrison}, F.~A., {et~al.} 2015,
  \apj, 802, 15, \dodoi{10.1088/0004-637X/802/1/15}

\bibitem[{{Helder} \& {Vink}(2008)}]{2008ApJ...686.1094H}
{Helder}, E.~A., \& {Vink}, J. 2008, \apj, 686, 1094, \dodoi{10.1086/591242}

\bibitem[{{Hughes} {et~al.}(2000){Hughes}, {Rakowski}, {Burrows}, \&
  {Slane}}]{2000ApJ...528L.109H}
{Hughes}, J.~P., {Rakowski}, C.~E., {Burrows}, D.~N., \& {Slane}, P.~O. 2000,
  \apjl, 528, L109, \dodoi{10.1086/312438}

\bibitem[{{Hwang} {et~al.}(2000){Hwang}, {Holt}, \&
  {Petre}}]{2000ApJ...537L.119H}
{Hwang}, U., {Holt}, S.~S., \& {Petre}, R. 2000, \apjl, 537, L119,
  \dodoi{10.1086/312776}

\bibitem[{{Hwang} \& {Laming}(2012)}]{2012ApJ...746..130H}
{Hwang}, U., \& {Laming}, J.~M. 2012, \apj, 746, 130,
  \dodoi{10.1088/0004-637X/746/2/130}

\bibitem[{{Hwang} {et~al.}(2004){Hwang}, {Laming}, {Badenes}, {Berendse},
  {Blondin}, {Cioffi}, {DeLaney}, {Dewey}, {Fesen}, {Flanagan}, {Fryer},
  {Ghavamian}, {Hughes}, {Morse}, {Plucinsky}, {Petre}, {Pohl}, {Rudnick},
  {Sankrit}, {Slane}, {Smith}, {Vink}, \& {Warren}}]{2004ApJ...615L.117H}
{Hwang}, U., {Laming}, J.~M., {Badenes}, C., {et~al.} 2004, \apjl, 615, L117,
  \dodoi{10.1086/426186}

\bibitem[{{Ichinohe} {et~al.}(2019){Ichinohe}, {Simionescu}, {Werner},
  {Fabian}, \& {Takahashi}}]{ichinohe19}
{Ichinohe}, Y., {Simionescu}, A., {Werner}, N., {Fabian}, A.~C., \&
  {Takahashi}, T. 2019, \mnras, 483, 1744, \dodoi{10.1093/mnras/sty3257}

\bibitem[{{Ichinohe} {et~al.}(2021){Ichinohe}, {Simionescu}, {Werner},
  {Markevitch}, \& {Wang}}]{ichinohe21}
{Ichinohe}, Y., {Simionescu}, A., {Werner}, N., {Markevitch}, M., \& {Wang},
  Q.~H.~S. 2021, \mnras, 504, 2800, \dodoi{10.1093/mnras/stab1060}

\bibitem[{{Ichinohe} {et~al.}(2017){Ichinohe}, {Simionescu}, {Werner}, \&
  {Takahashi}}]{ichinohe17}
{Ichinohe}, Y., {Simionescu}, A., {Werner}, N., \& {Takahashi}, T. 2017,
  \mnras, 467, 3662, \dodoi{10.1093/mnras/stx280}

\bibitem[{{Ichinohe} {et~al.}(2015){Ichinohe}, {Werner}, {Simionescu}, {Allen},
  {Canning}, {Ehlert}, {Mernier}, \& {Takahashi}}]{ichinohe15}
{Ichinohe}, Y., {Werner}, N., {Simionescu}, A., {et~al.} 2015, \mnras, 448,
  2971, \dodoi{10.1093/mnras/stv217}

\bibitem[{{Iwasaki} {et~al.}(2019){Iwasaki}, {Ichinohe}, \&
  {Uchiyama}}]{iwasaki19}
{Iwasaki}, H., {Ichinohe}, Y., \& {Uchiyama}, Y. 2019, \mnras, 488, 4106,
  \dodoi{10.1093/mnras/stz1990}

\bibitem[{Lee {et~al.}(1996)Lee, Wolberg, Chwa, \& Shin}]{lee96}
Lee, S., Wolberg, G., Chwa, K.-Y., \& Shin, S.~Y. 1996, IEEE Transactions on
  Visualization and Computer Graphics, 2, 337, \dodoi{10.1109/2945.556502}

\bibitem[{Lee {et~al.}(1997)Lee, Wolberg, \& Shin}]{lee97}
Lee, S., Wolberg, G., \& Shin, S. 1997, IEEE Transactions on Visualization and
  Computer Graphics, 3, 228, \dodoi{10.1109/2945.620490}

\bibitem[{Liu \& Nocedal(1989)}]{liu89}
Liu, D.~C., \& Nocedal, J. 1989, MATHEMATICAL PROGRAMMING, 45, 503

\bibitem[{Lowe(1999)}]{lowe99}
Lowe, D. 1999, in Proceedings of the Seventh IEEE International Conference on
  Computer Vision, Vol.~2, 1150--1157 vol.2, \dodoi{10.1109/ICCV.1999.790410}

\bibitem[{Lowekamp {et~al.}(2013)Lowekamp, Chen, Ibanez, \&
  Blezek}]{lowekamp13}
Lowekamp, B., Chen, D., Ibanez, L., \& Blezek, D. 2013, Frontiers in
  Neuroinformatics, 7, \dodoi{10.3389/fninf.2013.00045}

\bibitem[{Lucas \& Kanade(1981)}]{lucas81}
Lucas, B.~D., \& Kanade, T. 1981, in Proceedings of the 7th International Joint
  Conference on Artificial Intelligence - Volume 2, IJCAI'81 (San Francisco,
  CA, USA: Morgan Kaufmann Publishers Inc.), 674–679

\bibitem[{Macqueen(1967)}]{macqueen67}
Macqueen, J. 1967, Proceedings of the fifth Berkeley symposium on mathematical
  statistics and probability, 1967, 1, 281.
\newblock \url{https://ci.nii.ac.jp/naid/10019277817/en/}

\bibitem[{{Matsuda} {et~al.}(2020){Matsuda}, {Tanaka}, {Uchida}, {Amano}, \&
  {Tsuru}}]{2020PASJ...72...85M}
{Matsuda}, M., {Tanaka}, T., {Uchida}, H., {Amano}, Y., \& {Tsuru}, T.~G. 2020,
  \pasj, 72, 85, \dodoi{10.1093/pasj/psaa075}

\bibitem[{{Matsuda} {et~al.}(2022){Matsuda}, {Uchida}, {Tanaka}, {Yamaguchi},
  \& {Tsuru}}]{2022ApJ...940..105M}
{Matsuda}, M., {Uchida}, H., {Tanaka}, T., {Yamaguchi}, H., \& {Tsuru}, T.~G.
  2022, \apj, 940, 105, \dodoi{10.3847/1538-4357/ac94cf}

\bibitem[{Mattes {et~al.}(2003)Mattes, Haynor, Vesselle, Lewellen, \&
  Eubank}]{mattes03}
Mattes, D., Haynor, D., Vesselle, H., Lewellen, T., \& Eubank, W. 2003, IEEE
  Transactions on Medical Imaging, 22, 120, \dodoi{10.1109/TMI.2003.809072}

\bibitem[{{Orlando} {et~al.}(2022){Orlando}, {Wongwathanarat}, {Janka},
  {Miceli}, {Nagataki}, {Ono}, {Bocchino}, {Vink}, {Milisavljevic}, {Patnaude},
  \& {Peres}}]{2022A&A...666A...2O}
{Orlando}, S., {Wongwathanarat}, A., {Janka}, H.~T., {et~al.} 2022, \aap, 666,
  A2, \dodoi{10.1051/0004-6361/202243258}

\bibitem[{{Patnaude} \& {Fesen}(2009)}]{2009ApJ...697..535P}
{Patnaude}, D.~J., \& {Fesen}, R.~A. 2009, \apj, 697, 535,
  \dodoi{10.1088/0004-637X/697/1/535}

\bibitem[{{Patnaude} \& {Fesen}(2014)}]{2014ApJ...789..138P}
---. 2014, \apj, 789, 138, \dodoi{10.1088/0004-637X/789/2/138}

\bibitem[{{Patnaude} {et~al.}(2011){Patnaude}, {Vink}, {Laming}, \&
  {Fesen}}]{2011ApJ...729L..28P}
{Patnaude}, D.~J., {Vink}, J., {Laming}, J.~M., \& {Fesen}, R.~A. 2011, \apjl,
  729, L28, \dodoi{10.1088/2041-8205/729/2/L28}

\bibitem[{Pelleg \& Moore(2000)}]{10.5555/645529.657808}
Pelleg, D., \& Moore, A.~W. 2000, in Proceedings of the Seventeenth
  International Conference on Machine Learning, ICML '00 (San Francisco, CA,
  USA: Morgan Kaufmann Publishers Inc.), 727–734

\bibitem[{Rousseeuw(1987)}]{ROUSSEEUW198753}
Rousseeuw, P.~J. 1987, Journal of Computational and Applied Mathematics, 20,
  53, \dodoi{https://doi.org/10.1016/0377-0427(87)90125-7}

\bibitem[{Rublee {et~al.}(2011)Rublee, Rabaud, Konolige, \& Bradski}]{rublee11}
Rublee, E., Rabaud, V., Konolige, K., \& Bradski, G. 2011, in 2011
  International Conference on Computer Vision, 2564--2571,
  \dodoi{10.1109/ICCV.2011.6126544}

\bibitem[{Rueckert {et~al.}(1999)Rueckert, Sonoda, Hayes, Hill, Leach, \&
  Hawkes}]{rueckert99}
Rueckert, D., Sonoda, L., Hayes, C., {et~al.} 1999, IEEE Transactions on
  Medical Imaging, 18, 712, \dodoi{10.1109/42.796284}

\bibitem[{{Sanders}(2006)}]{sanders06}
{Sanders}, J.~S. 2006, \mnras, 371, 829,
  \dodoi{10.1111/j.1365-2966.2006.10716.x}

\bibitem[{{Sato} {et~al.}(2018){Sato}, {Katsuda}, {Morii}, {Bamba}, {Hughes},
  {Maeda}, {Ishida}, \& {Fraschetti}}]{2018ApJ...853...46S}
{Sato}, T., {Katsuda}, S., {Morii}, M., {et~al.} 2018, \apj, 853, 46,
  \dodoi{10.3847/1538-4357/aaa021}

\bibitem[{{Sato} {et~al.}(2017){Sato}, {Maeda}, {Bamba}, {Katsuda}, {Ohira},
  {Yamazaki}, {Masai}, {Matsumoto}, {Sawada}, {Terada}, {Hughes}, \&
  {Ishida}}]{2017ApJ...836..225S}
{Sato}, T., {Maeda}, Y., {Bamba}, A., {et~al.} 2017, \apj, 836, 225,
  \dodoi{10.3847/1538-4357/836/2/225}

\bibitem[{Schwarz(1978)}]{10.1214/aos/1176344136}
Schwarz, G. 1978, The Annals of Statistics, 6, 461 ,
  \dodoi{10.1214/aos/1176344136}

\bibitem[{{Stage} {et~al.}(2006){Stage}, {Allen}, {Houck}, \&
  {Davis}}]{2006NatPh...2..614S}
{Stage}, M.~D., {Allen}, G.~E., {Houck}, J.~C., \& {Davis}, J.~E. 2006, Nature
  Physics, 2, 614, \dodoi{10.1038/nphys391}

\bibitem[{{Tsuji} {et~al.}(2021){Tsuji}, {Uchiyama}, {Khangulyan}, \&
  {Aharonian}}]{2021ApJ...907..117T}
{Tsuji}, N., {Uchiyama}, Y., {Khangulyan}, D., \& {Aharonian}, F. 2021, \apj,
  907, 117, \dodoi{10.3847/1538-4357/abce65}

\bibitem[{{Uchiyama} \& {Aharonian}(2008)}]{2008ApJ...677L.105U}
{Uchiyama}, Y., \& {Aharonian}, F.~A. 2008, \apjl, 677, L105,
  \dodoi{10.1086/588190}

\bibitem[{{Uchiyama} {et~al.}(2007){Uchiyama}, {Aharonian}, {Tanaka},
  {Takahashi}, \& {Maeda}}]{2007Natur.449..576U}
{Uchiyama}, Y., {Aharonian}, F.~A., {Tanaka}, T., {Takahashi}, T., \& {Maeda},
  Y. 2007, \nat, 449, 576, \dodoi{10.1038/nature06210}

\bibitem[{{Vink} {et~al.}(2022){Vink}, {Patnaude}, \&
  {Castro}}]{2022ApJ...929...57V}
{Vink}, J., {Patnaude}, D.~J., \& {Castro}, D. 2022, \apj, 929, 57,
  \dodoi{10.3847/1538-4357/ac590f}

\bibitem[{Yaniv {et~al.}(2018)Yaniv, Lowekamp, Johnson, \& Beare}]{yaniv18}
Yaniv, Z., Lowekamp, B.~C., Johnson, H.~J., \& Beare, R. 2018, Journal of
  Digital Imaging, 31, 290, \dodoi{10.1007/s10278-017-0037-8}

\bibitem[{Zhu {et~al.}(1997)Zhu, Byrd, Lu, \& Nocedal}]{zhu97}
Zhu, C., Byrd, R.~H., Lu, P., \& Nocedal, J. 1997, ACM Trans. Math. Softw., 23,
  550–560, \dodoi{10.1145/279232.279236}

\end{thebibliography}
\bibliographystyle{aasjournal}

\end{document}